\documentclass[a4paper,12pt]{article}
\usepackage{amsmath, amssymb, latexsym, amscd, amsthm,amsfonts,amstext}
\usepackage[mathscr]{eucal}
\usepackage{graphicx}
\usepackage{subfig}

\numberwithin{equation}{section}
\input{tcilatex}
\begin{document}

\title{Uniqueness of a 3-D coefficient inverse scattering problem without
the phase information}
\author{Michael V. Klibanov$^{\ast }$ and Vladimir G. Romanov$^{\ast \ast }$
\and $^{\ast \ast }$Department of Mathematics and Statistics, University of
North \and Carolina at Charlotte, Charlotte, NC 28223, USA \and $^{\ast \ast
}$Sobolev Institute of Mathematics, Novosibirsk 630090, \and Russian
Federation \and \texttt{mklibanv@uncc.edu,romanov{@}math.nsc.ru}. }
\date{}
\maketitle

\begin{abstract}
We use a new method to prove uniqueness theorem for a coefficient inverse
scattering problem without the phase information for the 3-D Helmholtz
equation. We consider the case when only the modulus of the scattered wave
field is measured and the phase is not measured. The spatially distributed
refractive index is the subject of the interest in this problem.
Applications of this problem are in imaging of nanostructures and biological
cells.
\end{abstract}

\graphicspath{
{FIGURES/}
 {pics/}}



\section{Introduction}

\label{sec:1}

We consider an inverse problem of the determination of an unknown
coefficient of the 3-D Helmholtz equation from measurements of only the
modulus of the scattering part of the solution of this equation outside the
scatterer. Since the phase of the complex valued wave field is not measured
and since we search for an unknown coefficient of the 3-D Helmholtz
equation, we call our problem Coefficient Phaseless Inverse Scattering
Problem (CPISP). The goal of this paper is to prove uniqueness theorem for
this problem. The method of our proof is new. It was not used in proofs of
previous uniqueness results for CPISPs \cite%
{KS,KSIAP,AML,AA,KlibHelm,KlibHelm1}. Recall that it is assumed in the
majority of publications about inverse scattering problems that both the
phase and the modulus of the complex valued wave field are measured, see,
e.g. \cite{Bao3,Hu,Is,Li,Nov1,Nov2}.

Let $u_{0}$ be the incident wave field generated by a point source. Let $%
u_{sc}$ be the wave field, which occurs due to scattering by the scatterer.
The total wave field is $u=u_{0}+u_{sc}$. The goal of this paper is to prove
uniqueness theorem for a CPISP in the case when the modulus $\left\vert
u_{sc}\right\vert $ of the scattered wave field is measured on a certain
surface. In the previous publication \cite{KlibHelm} uniqueness was proven
for the case when the modulus $\left\vert u\right\vert $ of the total wave
field is measured. Compared with \cite{KlibHelm}, the main difficulty here
is caused by the interference of two wave fields: the total wave field and
the incident wave field. In \cite{KlibHelm1} uniqueness was also proven for
the case when $\left\vert u_{sc}\right\vert ^{2}$ is measured. However,
there is one inconvenient condition of the uniqueness theorem of \cite%
{KlibHelm1}, see Remark in section 2. Using the above mentioned new idea, we
lift this condition here.

CPISPs have applications in imaging of nanostructures whose sizes are about
100 nanometers, which is 0.1 micron. Hence, the wavelength of the probing
radiation must be also about 0.1 micron ($0.1\mu m)$. In this case the
frequencies are millions of gigahertz \cite{photonics}. It is well known
that it is possible to measure only the intensity of the scattered wave at
such huge frequencies, whereas the phase cannot be measured \cite%
{Dar,Pet,Ruhl}. The intensity is the square of the modulus. To image a
nanostructure, one needs to compute its unknown spatially distributed
dielectric constant using measurements of only the intensity of the
scattered wave field. Also, CPISPs have applications in optical imaging of
biological cells, since their sizes are between $1\mu m$ and $10\mu m$ \cite%
{Phillips}.

The light generated by lasers is used in this imaging. The diameter of the
laser beam is a few millimeters ($mm$). Recall that $1mm=10^{3}\mu m.$
Hence, inside of the laser beam the intensity of this beam significantly
exceeds the intensity of the portion of light scattered by nanostructures.
In addition, it is well known that if a light detector is placed inside of
this beam, then it is burned. Therefore, these detectors are placed outside
of this beam. This means that so placed detectors measure only the intensity
of the portion of light, which is scattered by those nanostructures, i.e.
they measure $\left\vert u_{sc}\right\vert ^{2}.$ We point out that the
precise mathematical model of the laser beam is outside of the scope of this
publication. So, the above was given only to explain why it is important to
consider the case when the function $\left\vert u_{sc}\right\vert ^{2}$
rather than the function $\left\vert u\right\vert ^{2}$ is assumed to be
given outside of a scatterer. \textbf{\ }

For the first time, the uniqueness result for a CPISP was proven in \cite{KS}%
. This was done for the 1-D case. As to the 3-D case, first uniqueness
theorems were proven in \cite{KSIAP,AML} for the case when the Schr\"{o}%
dinger equation 
\begin{equation}
\Delta v+k^{2}v-q\left( x\right) v=-\delta \left( x-y\right) ,x\in \mathbb{R}%
^{3}  \label{1.1}
\end{equation}%
is the underlying one and the potential $q\left( x\right) $ is unknown. We
note that equation (\ref{1.1}) is easier to work with than with the
Helmholtz equation. This is because, unlike the Helmholtz equation, the
potential $q\left( x\right) $ is not multiplied by $k^{2}$ in (\ref{1.1}).
The multiplication of the unknown coefficient by $k^{2}$ prompts the use of
the apparatus of the Riemannian geometry in the case of the Helmholtz
equation, see \cite{KlibHelm,KlibHelm1}, which is unlike (\ref{1.1}). In
addition to uniqueness theorems, reconstruction procedures for 3-D CPISPs
were developed by the authors both for the Schr\"{o}dinger equation \cite%
{KR,KR3} and for the Helmholtz equation \cite{KR1,KR2}. A modified
reconstruction procedure of \cite{KR1} was numerically implemented in \cite%
{KNP}.

Our CPISP is over-determined: the data depend on five variables, whereas the
unknown coefficient depends on three variables. On the other hand, the
authors are unaware about uniqueness results for 3-D Coefficient Inverse
Scattering Problems which would not use over-determined data even in the
case when both the phase and the modulus of the scattered waves are
measured. As to the uniqueness theorems for non over-determined 3-D
coefficient inverse problems without the phase information, we refer to \cite%
{AA} for the case of the Helmholtz equation with single measurement data.
The price to pay for this is the assumption that the right hand side of the
Helmholtz equation is a non-vanishing function $r\left( x\right) ,$ whereas
the right hand side in \cite{KlibHelm,KlibHelm1} and the current paper is
the $\delta -$function. In addition, similar results for the Schr\"{o}dinger
equation are in theorems 3, 4 of \cite{KSIAP} and in theorem 2 of \cite{AML}%
. In all these latter results for single measurement data the method of \cite%
{BukhKlib} is applied on the last step of the proof. This method is based on
Carleman estimates, also see, e.g. the recent survey of this method in \cite%
{Ksurvey}.

CPISPs were also considered by Novikov in \cite{Nov3,Nov4}. Statements of
CPISPs in \cite{Nov3,Nov4} differ from ours in some respects. In these
publications, uniqueness theorems are proven and reconstruction procedures
are developed.

Recall that a CPISP is about the reconstruction of an unknown coefficient
from phaseless measurements. Along with reconstructions of unknown
coefficients in CPISPs, phaseless inverse problems of the reconstruction of
unknown surfaces of scatterers are also attractive. In this regard, we refer
to \cite{Am,Bao1,Bao2,Iv1,Iv2,Li1} for numerical solutions of some inverse
scattering problems without the phase information in the case when the
surface of a scatterer was reconstructed. In addition, in \cite{Wang} the
phaseless inverse problem of the reconstruction of a source was considered.

In section 2 we formulate our CPISP as well as the uniqueness theorem of our
paper. In section 3 prove this theorem.

\section{Statement of the Problem}

\label{sec:2}

Below $x=\left( x_{1},x_{2},x_{3}\right) \in \mathbb{R}^{3}.$ Consider a
non-magnetic and non-conductive medium, which occupies the entire space $%
\mathbb{R}^{3}.$ Let $\Omega $ $\subset \mathbb{R}^{3}$ be a bounded domain.
Let $S\in C^{2}$ be a surface, which is the boundary of another bounded
domain $G\subset \mathbb{R}^{3},S=\partial G.$ We assume that $\Omega $ $%
\subseteq G$. Hence, $S\cap \Omega =\varnothing ,$ although the surface $S$
might be the boundary of $\Omega .$ Let $n^{2}(x)$ be the spatially varying
dielectric constant of the medium, where $n(x)$ is the refractive index. We
assume below that the function $n(x)$ satisfies the following conditions: 
\begin{equation}
n(x)\in C^{15}(\mathbb{R}^{3}),  \label{2.101}
\end{equation}%
\begin{equation}
n(x)\geq 1\text{ in }\mathbb{R}^{3},  \label{2.102}
\end{equation}%
\begin{equation}
n(x)=1\quad \text{for }\>x\in \mathbb{R}^{3}\setminus \Omega .  \label{2.103}
\end{equation}%
Condition (\ref{2.102}) means that the refractive index of the medium is not
less than the refractive index of the vacuum, which equals 1. Condition (\ref%
{2.103}) means that vacuum is outside of the domain $\Omega $. To explain
the smoothness condition (\ref{2.101}), we note that Lemma 1 formulated
below follows from \ results of \cite{KR1,KlibHelm}. On the other hand,
those results of \cite{KR1} use the fundamental solution of the hyperbolic
equation 
\begin{equation}
n^{2}(x)v_{tt}=\Delta v.  \label{2.1030}
\end{equation}
The construction of this solution works only if $n(x)\in C^{15}(\mathbb{R}%
^{3})$ \cite{KR1,R3}. In addition, the constructions of \cite{KR1,R3}
require the regularity of geodesic lines, see Condition below. We also note
that the minimal smoothness requirements for unknown coefficients are rarely
a significant concern in uniqueness theorems for multidimensional
coefficient inverse problems, see, e.g. \cite{Nov1,Nov2}, theorem 4.1 in
Chapter 4 of \cite{R2} and \cite{Ksurvey}.

The function $n(x)$ generates the conformal Riemannian metric, 
\begin{equation}
d\tau =n(x)\left\vert dx\right\vert ,|dx|=\sqrt{%
(dx_{1})^{2}+(dx_{2})^{2}+(dx_{3})^{2}}.  \label{2.105}
\end{equation}%
\ Everywhere below we assume without further mentioning that the following
condition holds:

\textbf{Condition}. \emph{Geodesic lines generated by the metric (\ref{2.105}%
) are regular. In other words, each pair of points }$x,y\in \mathbb{R}^{3}$%
\emph{\ can be connected by a single geodesic line }$\Gamma \left(
x,y\right) $\emph{.}

A sufficient condition of the regularity of geodesic lines is \cite{R4}%
\begin{equation*}
\sum_{i,j=1}^{3}\frac{\partial ^{2}\ln n(x)}{\partial x_{i}\partial x_{j}}%
\xi _{i}\xi _{j}\geq 0,\>\forall \xi \in \mathbb{R}^{3},\forall x\in 
\overline{\Omega }.
\end{equation*}%
For an arbitrary pair of points $x,y\in \mathbb{R}^{3}$ consider the travel
time $\tau (x,y)$ between them due to the Riemannian metric (\ref{2.105}).
Then \cite{R2} 
\begin{equation}
\tau \left( x,y\right) =\dint\limits_{\Gamma \left( x,y\right) }n\left( \xi
\right) d\sigma ,  \label{2.106}
\end{equation}%
where $d\sigma $ is the euclidean arc length.

Let $y\in \mathbb{R}^{3}$ be the position of the point source, $r=\left\vert
x-y\right\vert $ and $k>0$ be the wavenumber. We consider the Helmholtz
equation with the radiation condition at the infinity 
\begin{equation}
\Delta u+k^{2}n^{2}(x)u=-\delta (x-y),\quad x\in \mathbb{R}^{3},
\label{2.109}
\end{equation}%
\begin{equation}
\partial _{r}u-iku=o\left( 1/r\right) ,\>r\rightarrow \infty .  \label{2.110}
\end{equation}%
Let $u_{0}$ be the incident spherical wave and $u_{sc}$ be the scattered
wave,%
\begin{equation}
u_{0}\left( x,y,k\right) =A_{0}(x,y)e^{ik\left\vert x-y\right\vert },\quad
A_{0}(x,y)=\frac{1}{4\pi \left\vert x-y\right\vert },  \label{2.1100}
\end{equation}%
\begin{equation}
u_{sc}\left( x,y,k\right) =u\left( x,y,k\right) -u_{0}\left( x,y,k\right) .
\label{2.111}
\end{equation}

We model the propagation of the electric wave field in $\mathbb{R}^{3}$ by
the solution of the problem (\ref{2.109}), (\ref{2.110}). This model was
justified numerically in \cite{BMM} in the case of the time domain.
Numerical results of section 7.2.2 of \cite{BMM} demonstrate that this model
can replace the modeling via the full Maxwell's system, provided that only a
single component of the electric field is incident upon the medium. Then
this component dominates two other components while propagating through the
medium. Furthermore, the propagation of this component is well governed by
the single PDE (\ref{2.1030}), which is the time domain analog of equation (%
\ref{2.109}), see Figure 19b in \cite{BMM} and the discussion in the
paragraph just above section 8 of \cite{BMM}. This conclusion was verified
via accurate imaging using electromagnetic experimental data in, e.g.
Chapter 5 of \cite{BK} and \cite{TBKF1,TBKF2}.

Let $\left( a,b\right) \subset \left\{ k:k>0\right\} $\ be an arbitrary
interval.\emph{\ }Our interest in this paper is in the following CPISP:

\textbf{Coefficient Phaseless Inverse Scattering Problem (CPISP)}. \emph{%
Suppose that the function }$n\left( x\right) $\emph{\ satisfies conditions (%
\ref{2.101})-(\ref{2.103}). Assume that the following function }$f\left(
x,y,k\right) $\emph{\ is given}%
\begin{equation}
f\left( x,y,k\right) =\left\vert u_{sc}\left( x,y,k\right) \right\vert ^{2},
\>\forall x,y\in S,x\neq y, \forall k\in \left( a,b\right).  \label{2.112}
\end{equation}%
\emph{Determine the function }$n\left( x\right) $\emph{\ for }$x\in \Omega .$

For an arbitrary number $\theta >0$ denote $\mathbb{C}_{\theta }=\left\{
z\in \mathbb{C}:\func{Im}z>-\theta \right\} .$

\textbf{Lemma 1}. \emph{Choose an arbitrary bounded domain }$G_{1}\subset 
\mathbb{R}^{3}$\emph{\ such that }$G\subset G_{1}$\emph{\ and }$S\cap
\partial G_{1}=\varnothing .$\emph{\ Then there exists a number }$\theta
=\theta \left( G_{1}\right) >0$\emph{\ such that for each }$y\in \mathbb{R}%
^{3}$\emph{\ and for all }$x\in G_{1},x\neq y$\emph{\ the function }$%
u_{sc}\left( x,y,k\right) $\emph{\ is analytic as the function of }$k\in 
\mathbb{C}_{\theta }.$\emph{\ In particular, this implies that the function }%
$f\left( x,y,k\right) $\emph{\ is uniquely determined for all }$k>0$\emph{\
by its values for }$k\in \left( a,b\right) .$\emph{\ Next, for any pair }$%
x,y\in \mathbb{R}^{3},x\neq y$\emph{\ the asymptotic behavior of the
function }$u_{sc}(x,y,k)$\emph{\ as} $k\rightarrow \infty $ \emph{is}%
\begin{equation}
u_{sc}(x,y,k)=A(x,y)e^{ik\tau \left( x,y\right) }-A_{0}(x,y)e^{ik\left\vert
x-y\right\vert }+\widehat{u}(x,y,k),  \label{3.1}
\end{equation}%
\emph{where the function }$\widehat{u}(x,y,k)=O\left( 1/k\right) $ \emph{as} 
$k\rightarrow \infty $ \emph{and the function }$A(x,y)>0.$ \emph{%
Furthermore, the function }$\widehat{u}(x,y,k)$\emph{\ is such that } 
\begin{equation}
\frac{\partial }{\partial k}\emph{\ }\widehat{u}(x,y,k)=O\left( \frac{1}{k}%
\right) ,\>k\rightarrow \infty .  \label{3.20}
\end{equation}

Lemma 1 follows from results of \cite{KR1,KlibHelm}. We note that the
analyticity of the function $u_{sc}\left( x,y,k\right) $ with respect $k$
was proven in \cite{KlibHelm} using results of the book \cite{V}. Theorem 1
is the main result of this paper.

\textbf{Theorem 1}. \emph{For any fixed pair of points }$x,y\in S,x\neq y$%
\emph{\ the number }$\tau \left( x,y\right) $\emph{\ is uniquely determined
from the function }$f\left( x,y,k\right) .$\emph{\ Also,} \emph{there exists
at most one solution of the CPISP.}

We assume below that conditions (\ref{2.101})-(\ref{2.103}) hold true and
devote the rest of this paper to the proof of this theorem.

\textbf{Remark}. Since $S\cap \Omega =\varnothing ,$ then it follows from (%
\ref{2.103}) and (\ref{2.112}) that measurements are performed outside of
the domain $\Omega $ where possible heterogeneities are$.$ Thus, (\ref{2.112}%
) removes an inconvenient assumption of \cite{KlibHelm1}, which requires to
perform measurements at a surface, which is located inside of the domain
with heterogeneities.

\section{Proof}

\label{sec:3}

Fix a pair of points $x,y\in S,x\neq y$ and consider the asymptotic
expansion of the function $f\left( x,y,k\right) $ at $k\rightarrow \infty .$
Denote%
\begin{equation}
\alpha =\alpha \left( x,y\right) =\tau (x,y)-\left\vert x-y\right\vert .
\label{3.3}
\end{equation}%
By Lemma 1 the function $f\left( x,y,k\right) $ is known for all $k>0.$ It
follows from (\ref{2.112})-(\ref{3.3}) that%
\begin{equation}
f\left( x,y,k\right) =A^{2}(x,y)+A_{0}^{2}(x,y)-2A(x,y)A_{0}(x,y)\cos \left(
k\alpha \right) +\hat{f}\left( x,y,k\right) ,  \label{3.4}
\end{equation}%
where the functions $\hat{f}(x,y,k)=O\left( 1/k\right) $ and $\partial _{k}%
\hat{f}(x,y,k)=O\left( 1/k\right) $ as $k\rightarrow \infty $. We prove
below that the number $\alpha $ can be uniquely recovered from the function $%
f(x,y,k)$.

First, we find $A(x,y)$. To do this, we prove first that 
\begin{equation}
\lim_{k^{\prime }\rightarrow \infty }\sup_{k\in (k^{\prime },\infty )}\left[
-2A(x,y)A_{0}(x,y)\cos \left( k\alpha \right) +\hat{f}\left( x,y,k\right) %
\right] =2A(x,y)A_{0}(x,y).  \label{3.4a}
\end{equation}%
Indeed, since $A(x,y)>0$ and $A_{0}(x,y)>0$, we have%
\begin{equation*}
\sup_{k\in (k^{\prime },\infty )}\left[ 2A(x,y)A_{0}(x,y)\cos \left( k\alpha
\right) +\hat{f}\left( x,y,k\right) \right] \geq
\end{equation*}%
\begin{equation*}
\sup_{k\in (k^{\prime },\infty )}\left[ 2A(x,y)A_{0}(x,y)\cos \left( k\alpha
\right) \right] -\sup_{k\in (k^{\prime },\infty )}|\hat{f}\left(
x,y,k\right) |=
\end{equation*}%
\begin{equation*}
2A(x,y)A_{0}(x,y)-\sup_{k\in (k^{\prime },\infty )}|\hat{f}\left(
x,y,k\right) |.
\end{equation*}%
On the other hand,%
\begin{equation*}
\sup_{k\in (k^{\prime },\infty )}\left[ 2A(x,y)A_{0}(x,y)\cos \left( k\alpha
\right) +\hat{f}\left( x,y,k\right) \right] \leq
2A(x,y)A_{0}(x,y)+\sup_{k\in (k^{\prime },\infty )}|\hat{f}\left(
x,y,k\right) |.
\end{equation*}%
Hence, we have obtained that%
\begin{equation*}
2A(x,y)A_{0}(x,y)-\sup_{k\in (k^{\prime },\infty )}|\hat{f}\left(
x,y,k\right) |\leq
\end{equation*}%
\begin{equation}
\sup_{k\in (k^{\prime },\infty )}\left[ 2A(x,y)A_{0}(x,y)\cos \left( k\alpha
\right) +\hat{f}\left( x,y,k\right) \right] \leq  \label{3.40}
\end{equation}%
\begin{equation*}
2A(x,y)A_{0}(x,y)+\sup_{k\in (k^{\prime },\infty )}|\hat{f}\left(
x,y,k\right) |.
\end{equation*}
We also have $\sup_{k\in (k^{\prime },\infty )}|\hat{f}\left( x,y,k\right)
|=O\left( {1}/{k^{\prime }}\right) $. Hence, taking the limit in first and
third lines of (\ref{3.40}) as $k^{\prime }\rightarrow \infty ,$ we obtain (%
\ref{3.4a}). Therefore, 
\begin{equation}
f^{\ast }(x,y):=\lim_{k^{\prime }\rightarrow \infty }\sup_{k\in (k^{\prime
},\infty )}f\left( x,y,k\right) =\left( A(x,y)+A_{0}(x,y)\right) ^{2}.
\label{3.4aa}
\end{equation}%
\ The number $A(x,y)>0$ can be uniquely found from (\ref{3.4aa}) as 
\begin{equation*}
A(x,y)=\sqrt{f^{\ast }(x,y)}-A_{0}(x,y).
\end{equation*}

Introduce the functions $g$ and $p$, 
\begin{equation*}
g(x,y,k)=\frac{A^{2}(x,y)+A_{0}^{2}(x,y)-f\left( x,y,k\right) }{%
2A(x,y)A_{0}(x,y)},
\end{equation*}%
\begin{equation*}
p(x,y,k)=\frac{1}{2A(x,y)A_{0}(x,y)}\hat{f}(x,y,k).
\end{equation*}%
Then equation (\ref{3.4}) can be rewritten in the form 
\begin{equation}
g(x,y,k)=\cos (k\alpha )-p(x,y,k),  \label{3.5}
\end{equation}%
where the function $g(x,y,k)$ is known and $p(x,y,k)=O(1/k)$, $\partial
_{k}p(x,y,k)=O(1/k)$.

First, if $\alpha \neq 0,$ then it follows from (\ref{3.5}) that the limit 
\begin{equation}
\lim_{k\rightarrow \infty }g\left( x,y,k\right)  \label{3.6}
\end{equation}%
does not exist. On the other hand, if $\alpha =0,$ then $\lim_{k\rightarrow
\infty }g\left( x,y,k\right) =0.$ Thus, we have established that the limit (%
\ref{3.6}) exists if and only if (see (\ref{3.3}))%
\begin{equation}
\tau (x,y)=\left\vert x-y\right\vert .  \label{3.7}
\end{equation}%
Since the function $g$ is known, then one can establish whether or not the
limit (\ref{3.6}) exists, thus establishing whether or not (\ref{3.7}) holds.

Assume now that (\ref{3.7}) does not hold. Hence, $\alpha \neq 0.$ By (\ref%
{2.106}) and (\ref{3.3}) 
\begin{equation}
\alpha(x,y) >0.  \label{3.70}
\end{equation}%
We show now how to find the number $\alpha(x,y) $. First we show that there
exist a countable number of zeros $k_{n}$ of the function $g\left(
x,y,k\right) $ and 
\begin{equation}
\lim_{n\rightarrow \infty }k_{n}=\infty .  \label{3.10}
\end{equation}
We have: 
\begin{equation}
g\left( x,y,k\right) =0\Leftrightarrow \cos \left( k\alpha(x,y) \right)
=p\left( x,y,k\right) .  \label{3.11}
\end{equation}

Let 
\begin{equation}
k\alpha \in (\pi (n-1),\pi n)  \label{3.110}
\end{equation}
for a sufficiently large integer $n>1$. Then $p\left( x,y,k\right) =O(1/n)$
and also $\partial _{k}p\left( x,y,k\right) =O(1/n)$ as $n\rightarrow \infty 
$. Rewrite the equation $\cos (k\alpha )=p(x,y,k)$ in the form 
\begin{equation}
\sin (\pi /2+k\alpha )=-p(x,y,k).  \label{3.111}
\end{equation}
Since $\left\vert p(x,y,k)\right\vert <1$ for sufficiently large $n$ for $k$
satisfying (\ref{3.110}), then (\ref{3.111}) is equivalent with 
\begin{equation}
k\alpha =-\frac{\pi }{2}+(-1)^{(n+1)}\arcsin p\left( x,y,k\right) +n\pi .
\label{3.12}
\end{equation}%
For $k$ satisfying (\ref{3.110}), consider the function $F_{n}(k,x,y),$ 
\begin{equation*}
F_{n}(k,x,y)=k\alpha +\frac{\pi }{2}+(-1)^{(n)}\arcsin p\left( x,y,k\right)
-n\pi .
\end{equation*}%
Then 
\begin{equation*}
\partial _{k}F_{n}(x,y,k)=\alpha +O\left( \frac{1}{n}\right) ,n\rightarrow
\infty .
\end{equation*}%
Hence, $\partial _{k}F_{n}(x,y,k)>0$ for sufficiently large $n$. This means
that the function $F_{n}(x,y,k)$ is monotonically increasing with respect to 
$k$ on the interval (\ref{3.110}). Next, 
\begin{equation}
F_{n}(x,y,(n-1)\pi /\alpha )=-\frac{\pi }{2}+O\left( \frac{1}{n}\right)
,\quad F_{n}(x,y,n\pi /\alpha )=\frac{\pi }{2}+O\left( \frac{1}{n}\right) .
\label{3.12b}
\end{equation}%
Since the function $F_{n}(x,y,k)$ is continuous with respect to $k$ and has
different signs on the edges of the interval (\ref{3.110}), then the
monotonicity of this function implies that it has unique zero inside of this
interval. Denote this zero by $k_{n}$. The asymptotic formula for these
zeros follows from (\ref{3.12}): 
\begin{equation*}
k_{n}\alpha =-\frac{\pi }{2}+n\pi +O\left( \frac{1}{n}\right) ,\quad
n\rightarrow \infty .
\end{equation*}%
Hence,%
\begin{equation}
\alpha \left( k_{n+1}-k_{n}\right) =\pi +O\left( \frac{1}{n}\right) ,\quad
n\rightarrow \infty .  \label{3.14}
\end{equation}%
Therefore the number $\alpha $ is uniquely determined as%
\begin{equation}
\alpha =\lim_{n\rightarrow \infty }\frac{\pi }{k_{n+1}-k_{n}}.  \label{3.24}
\end{equation}

Finally, since we know the function $g\left( x,y,k\right) $ as the function
of $k$ and numbers $k_{n}$ are zeros of this function, then we know the
numbers $k_{n}$ as well. Hence, (\ref{3.3}) and (\ref{3.24}) imply that the
number $\tau \left( x,y\right) $ is determined uniquely from the function $%
f\left( x,y,k\right) $ in (\ref{2.112}) for any fixed pair of points $x,y\in
S,x\neq y.$ Therefore, the first assertion of Theorem 1 is proved.

We now prove the second assertion of Theorem 1. To do this, we apply theorem
3.4 of Chapter 3 of the book \cite{R2}. This is the theorem about stability
and uniqueness of the so-called Inverse Kinematic Problem of Seismic. We use
notations of that theorem for the convenience of the reader.

Suppose that there exist two coefficients $n_{1}\left( x\right) $ and $%
n_{2}\left( x\right) ,$ which generate the same function $f\left(
x,y,k\right) $ in (\ref{2.112}). Consider corresponding functions $\tau
_{1}\left( x,y\right) $ and $\tau _{2}\left( x,y\right) $ defined via (\ref%
{2.106}), 
\begin{equation*}
\tau _{1}\left( x,y\right) =\dint\limits_{\Gamma _{1}\left( x,y\right) }{%
n_{1}\left( \xi \right) }d\sigma ,\>\>\tau _{2}\left( x,y\right)
=\dint\limits_{\Gamma _{2}\left( x,y\right) }{n_{2}\left( \xi \right) }%
d\sigma ,\forall x,y\in \mathbb{R}^{3},
\end{equation*}%
where $\Gamma _{1}\left( x,y\right) $ and $\Gamma _{2}\left( x,y\right) $
are geodesic lines generated by functions $n_{1}\left( x\right) $ and $%
n_{2}\left( x\right) $ respectively. Then the first assertion of Theorem 1
implies that 
\begin{equation}
\tau _{1}\left( x,y\right) =\tau _{2}\left( x,y\right) ,\forall x,y\in S.
\label{3.26}
\end{equation}%
It follows from (\ref{2.102}) that 
\begin{equation}
n_{1}\left( x\right) ,n_{2}\left( x\right) \geq 1.  \label{3.27}
\end{equation}%
Using (\ref{2.101}) and (\ref{3.27}), we obtain that there exists a number $%
n_{00}>1$ such that 
\begin{equation}
\left\Vert n_{1}\left( x\right) \right\Vert _{C^{2}\left( \overline{G}%
\right) },\left\Vert n_{2}\left( x\right) \right\Vert _{C^{2}\left( 
\overline{G}\right) }\leq n_{00}.  \label{3.28}
\end{equation}%
Let $\Lambda \left( 1,n_{00}\right) $ be the set of functions $n\left(
x\right) $ defined in $\overline{G}$ and satisfying the following conditions:

\begin{enumerate}
\item The function $n\left( x\right) \in C^{15}\left( \overline{G}\right)
,\left\Vert n\left( x\right) \right\Vert _{C^{2}\left( \overline{G}\right)
}\leq n_{00}$ and also $n\left( x\right) \geq 1$ in $\overline{G}.$

\item The function $n\left( x\right) $ satisfies Condition of section 2
about the regularity of geodesic lines generated by metric (\ref{2.105}).
\end{enumerate}

By (\ref{3.26})-(\ref{3.28}) functions $n_{1}\left( x\right) ,n_{2}\left(
x\right) \in \Lambda \left( 1,n_{00}\right) .$ Thus, (\ref{3.26}) and the
estimate (3.66) of theorem 3.4 of Chapter 3 of the book \cite{R2} imply that 
$n_{1}\left( x\right) \equiv n_{2}\left( x\right) $. \ $\square $

\begin{center}
\textbf{Acknowledgments}
\end{center}

The work of MVK was supported by US Army Research Laboratory and US Army
Research Office grant W911NF-15-1-0233 and by the Office of Naval Research
grant N00014-15-1-2330. The work of VGR was partially supported by the
Russian Foundation for Basic Research grant No. 17-01-00120-a.

\end{document}